\documentstyle[aps,pre,twocolumn]{revtex}

\newcommand{\ptl}[2]{\frac{\partial #1}{\partial #2}}
\newcommand{\th}{\theta}
\newcommand{\et}{\eta}

\newcommand{\ett}{\tilde{\eta}}
\newcommand{\etb}{\bar{\eta}}
\newcommand{\rb}{\bar{\rho}}
\newcommand{\ep}{\epsilon}
\newcommand{\al}{\alpha}
\newcommand{\om}{\omega}
\newcommand{\lp}{{\cal L}_p}
\newcommand{\eb}{\bar{\ep}}

\begin{document}
\twocolumn[\hsize\textwidth\columnwidth\hsize\csname
@twocolumnfalse\endcsname

\draft

\title{Synchronization, chaos, and breakdown of collective domain
  oscillations in reaction-diffusion systems}

\author{C. B. Muratov} \address{Department of Physics, Boston
  University, Boston, Massachusetts, 02215}

\date{\today}

\maketitle

\begin{abstract}
  The universal equations describing collective oscillations of the
  multidomain patterns of small period in an arbitrary $d$-dimensional
  reaction-diffusion system of the activator-inhibitor type are
  asymptotically derived.  It is shown that not far from the instability
  leading to the formation of the pulsating multidomain pattern the
  oscillations of different domains synchronize. In one dimension
  standing and traveling waves of the oscillation phase are realized. In
  addition to these, in two dimensions target and spiral waves of the
  oscillation phase, as well as spatio-temporal chaos of domain
  oscillations, are feasible. Further inside the unstable region the
  collective oscillations break down, so the pulsating multidomain
  pattern transforms into an irregular pulsating pattern, the uniform
  self-oscillations, or turbulence. The parameter regions where these
  effects occur are analyzed. The effects of the pattern's disorder are
  also studied. The conclusions of the analysis are supported by the
  numerical simulations of a concrete model. The obtained results
  explain the dynamics of Turing patterns observed in the experiments on
  chlorite-iodide-malonic acid reaction.
\end{abstract}
\bibliographystyle{prsty}

\vskip2pc]

\section{introduction}

Complex dynamic patterns, such as traveling waves, breathers, or
spatio-temporal chaos, are encountered in the variety of nonequilibrium
physical, chemical, and biological systems
\cite{nicolis,cross93,field,mikhailov,kapral,murray,gurevich87,%
  ko:book,ko:ufn89,ko:ufn90,niedernostheide}. These systems include
electron-hole and gas plasma, semiconductor and superconductor
structures, systems with uniformly generated combustion material,
autocatalytic chemical reactions, models of population dynamics
\cite{cross93,field,mikhailov,kapral,murray,gurevich87,ko:book,%
  ko:ufn89,ko:ufn90,niedernostheide}.  Recently, an intriguing
phenomenon was observed in the chemical experiments with Turing patterns
\cite{boissande}, combustion system with cellular flames
\cite{gorman94:chaos}, and the catalytic reaction on the surface
\cite{rose96}. In these experiments the patterns that were observed
consisted of many disk-shaped domains whose radii oscillated in time.
The oscillations of different domains either synchronized or exhibited
complex spatio-temporal behavior.

Many nonequilibrium systems in which patterns can form, including the
systems mentioned above, are described by the reaction-diffusion systems
of the activator-inhibitor
type, the simplest of which is a pair of reaction-diffusion equations
\cite{cross93,field,mikhailov,kapral,murray,gurevich87,% 
  ko:book,ko:ufn89,ko:ufn90,niedernostheide}
\begin{equation} \label{1} 
  \tau_\th {\partial \th \over \partial t} = l^2 \Delta \th - q(\th,
  \et, A),
\end{equation} 
\begin{equation} \label{2} \tau_\et {\partial \et \over \partial
    t} = L^2 \Delta \et - Q(\th, \et, A),
\end{equation}
where $\th$ is the activator, $\et$ is the inhibitor, $l$ and $L$ are
the characteristic length scales and $\tau_\th$ and $\tau_\et$ are the
characteristic time scales of the activator and the inhibitor,
respectively; $q$ and $Q$ are certain non-linear functions, and $A$ is
the system's excitation level. For example, in the system with the
uniformly generated combustion material the activator is the temperature
of the gas mixture, the inhibitor is the density of the fuel, $A$ is
proportional to the rate of the fuel supply, and the non-linear
functions $q$ and $Q$ contain the dissipation, supply, and reaction
terms \cite{ko:book,ko:ufn89}. The length scales $l$ and $L$ are related
to the diffusion coefficients of the activator and the inhibitor $D_\th$
and $D_\et$, respectively: $l = \sqrt{D_\th \tau_\th}$ and $L =
\sqrt{D_\et \tau_\et}$.

Kerner and Osipov showed that the properties of the patterns forming in
the systems described by Eqs. (\ref{1}) and (\ref{2}) are determined
mainly by the parameters $\ep = l/L$ and $\al = \tau_\th / \tau_\et$ and
the form of the nullcline of Eq. (\ref{1}). For many systems this
nullcline is N-shaped (Fig. \ref{null})
\cite{ko:book,ko:ufn89,ko:ufn90}. In such N systems static domain
patterns form when $\ep \ll 1$ and $\al \gg \ep$ (KN systems), traveling
waves (autowaves) at $\al \ll 1$ and $\al \lesssim \ep^2$ ($\Omega$N
systems), and all sorts of dynamic patterns at $\ep \ll 1$ and $\ep^2
\lesssim \al \lesssim \ep$ (K$\Omega$N systems)
\cite{ko:book,ko:ufn89,ko:ufn90,mo1:pre96,mo2:pre96}. As a result of the
instability of the homogeneous state periodic and more complex patterns
form in these system, whereas when the homogeneous state of the system
is stable one can excite solitary patterns --- autosolitons (AS) --- by
a sufficiently strong localized external stimulus. In KN systems these
patterns are collections of static domains with sharp walls (interfaces)
whose width is of order $l$ \cite{ko:book,ko:ufn89,mo1:pre96,mo2:pre96}.
These domain patterns may undergo different kinds of instabilities
leading to the formation of complex dynamic patterns when $\al$ becomes
sufficiently small. The simplest example of such destabilization is the
transformation of a static AS to a pulsating (breathing) AS. This effect
was discovered by Koga and Kuramoto in an axiomatic reaction-diffusion
model \cite{koga80} and subsequently studied by many authors, both for
one-dimensional and higher-dimensional radially-symmetric AS
\cite{ko:book,ko:ufn89,ko:ufn90,mo1:pre96,ohta89,hagberg96}.  Pulsating
AS were directly observed in semiconductors \cite{mayer:pd88}, composite
superconductors \cite{baev82}, autocatalytic reaction \cite{haim96}, and
combustion experiments \cite{gorman94:extinct}.

The situation becomes much more complicated when instead of a single AS
the pattern consists of many interacting domains. Several attempts to
approach this problem were made for one-dimensional N systems. Kerner
and Osipov showed that the stationary periodic domain patterns (strata)
may undergo instability and transform into a breathing pattern
\cite{ko:book,ko:ufn90}. Ohta {\em et al.} were able to obtain the
linearized equation of motion for the periodic patterns in the
piecewise-linear reaction-diffusion model and showed that the
introduction of a simple nonlinearity to these equations leads to the
synchronization of the domain oscillations \cite{ohta90,ohta92}.  Still,
the question of the effect of the interaction of domain patterns
undergoing breathing motion, especially in higher-dimensional systems,
remains largely unresolved.

It is easy to show that in the case $\ep \ll 1$ the wave length of the
fluctuation with respect to which the Turing instability of the
homogeneous state of the system (\ref{1}) and (\ref{2}) is realized is
$\lambda \sim (l L)^{1/2}$ \cite{ko:book,ko:ufn90}. For this reason the
periodic Turing structures that form in the system have the period $\lp
\sim \ep^{1/2}L \ll L$.  Moreover, in the higher-dimensional systems
with $\ep \ll 1$ any static pattern will consist of the domains whose
characteristic size is of order $\ep^{1/3} L \ll L$
\cite{mo1:pre96,mo2:pre96,m:pre96}.  This means that in a pattern
consisting of many domains one domain interacts with a large number of
other domains at the same time. This fact should significantly reduce
the complexity of the interactions between different domains.

In this paper we will study the periodic one-dimensional and
two-dimensional hexagonal patterns of small period ($\lp \ll L$)
undergoing the oscillatory instability in an arbitrary K$\Omega$N
system.  Using the interfacial dynamics approach, we will derive the
universal nonlinear equations describing the pulsations of the periodic
multidomain patterns in arbitrary dimensions. We will analyze the
conditions for the synchronization and breakdown of the pulsations, the
effects of the disorder, and study possible complex spatio-temporal
behaviors. Finally, we will compare the effects studied by us with
relevant experiments.

Our paper is organized as follows. In Sec. II we reduce Eqs. (\ref{1})
and (\ref{2}) to the problem of interfacial dynamics in the limit $\ep
\rightarrow 0$ and show the way of treating this problem in the case
$\lp \ll 1$. In Sec. III we apply this method to the one-dimensional
periodic strata. In Sec. IV we consider two-dimensional hexagonal
patterns and also briefly discuss the case of the three-dimensional
patterns. In Sec. V we analyze the domain oscillations in a disordered
pattern with a random distribution of domain sizes. In Sec. VI we
present the results of the numerical simulations for a concrete model,
and, finally, in Sec. VII we discuss the relevancy of the obtained
results to the experiments and draw conclusions.

\section{interfacial dynamics problem} 

Let us measure the length and time in the units of $L$ and $\tau_\et$,
respectively. Then Eqs. (\ref{1}) and (\ref{2}) become
\begin{equation} \label{act}
  \al \ptl{\th}{t} = \ep^2 \Delta \th - q(\th, \et, A),
\end{equation}
\begin{equation} \label{inh}
  \ptl{\et}{t} = \Delta \et - Q(\th, \et, A).
\end{equation}
The boundary conditions for these equations may be neutral or periodic. 
From the mathematical point of view the fact that $\th$ is the activator
and $\et$ is the inhibitor means that for some values of $\th$ and $\et$
we have $q'_\th < 0$, and that for all $\th$ and $\et$
\cite{ko:book,ko:ufn89,ko:ufn90,mo1:pre96}
\begin{equation} \label{actinh}
  Q'_\et > 0, ~~~q'_\et Q'_\th < 0,
\end{equation}
and the derivatives $Q'_\et, Q'_\th$, and $q'_\et$ do not change sign.

The dynamics of the domain patterns forming in KN and K$\Omega$N systems
can be reduced to the interfacial dynamics problem in which the dynamics
of the pattern interfaces is coupled to the bulk field
\cite{ohta89,m:pre96}.  Far enough from the domain interfaces (at
distances much greater than $\ep$) the bulk field must satisfy the
equation of smooth distributions (outer solution)
\begin{equation} \label{sm}
  \ptl{\et}{t} = \Delta \et - Q \bigl( \th(\et), \et, A \bigr),
\end{equation}
in which $\th$ and $\et$ are related by the equation of local coupling
\begin{equation} \label{lc}
  q(\th, \et, A) = 0,
\end{equation}
that is, far from the domain interfaces $\th$ and $\et$ lie on the
nullcline of Eq. (\ref{act}). This relation is multivalued (see Fig.
\ref{null}), so one has to choose the branch with $\th < \th_0$ in the
``cold'' regions (the domains of low values of $\th$) and $\th > \th_0'$
in the ``hot'' regions (the domains of high values of $\th$). The
dynamics of the interface is governed by the following equation
\begin{equation} \label{v} 
  \vec{n} \cdot \ptl{\vec{r}}{t} = - \ep^2 \al^{-1} K(\vec{r}) +
  v(\et(\vec{r})),
\end{equation}
where $\vec{r}$ is a point on the interface, $\vec{n}$ is the normal
vector to the interface pointing into the cold region, $K(\vec{r})$ is
the curvature at the point $\vec{r}$ of the interface, and
$v(\et(\vec{r}))$ is the velocity of the interface in the absence of the
curvature as a function of the value of the bulk field $\et(\vec{r})$ at
the interface. The function $v(\et)$ is a solution to the nonlinear
eigenvalue problem and is in general some complicated nonlinear function
of $\et$ \cite{ohta89,m:pre96}.

The interfacial dynamics problem represented by Eqs. (\ref{sm}) and
(\ref{v}) is a highly nonlinear problem. However, the situation becomes
simpler if the pattern consists of the alternating hot and cold regions
whose characteristic size is much smaller than 1. Kerner and Osipov
showed that in this case the value of $\et$ is close to a constant in
the entire pattern \cite{ko:book,ko:ufn90}. The reason is that because
of its high diffusivity the inhibitor cannot react well on these
variations of the activator whose characteristic length is much smaller
than the characteristic length scale of the inhibitor. This allows us to
linearize Eq. (\ref{sm}) around $\et = \et_s$, where $\et_s$ is the
constant value of the inhibitor. Introducing
\begin{equation} \label{det}
  \ett = \et - \et_s,
\end{equation}
we get
\begin{equation} \label{smlin}
  \ptl{\ett}{t} = \Delta \ett - c_1 \ett - (c_3 - c_1) \ett I(x) -
  Q(\th_{s1}, \et_s) - a I(x),
\end{equation}
where
\begin{equation} \label{c13}
  c_{1,3} = Q'_\et(\th_{s1,3}, \et_s) - {q'_\et(\th_{s1,3}, \et_s)
    Q'_\th(\th_{s1,3}, \et_s) \over q'_\th(\th_{s1,3}, \et_s)},
\end{equation}
\begin{equation} \label{a}
  a = Q(\th_{s3}, \et_s) - Q(\th_{s1}, \et_s),
\end{equation}
the values of $\th_{s1,3}$ satisfy
\begin{equation}
  \label{ths13}
  q(\th_{s1,3}, \et_s) = 0,
\end{equation}
$\th_{s1}$ and $\th_{s3}$ being the minimal and the maximal roots of Eq.
(\ref{ths13}), respectively; $I(x)$ is the indicator function which is
equal to 1 if $x$ is in the hot region and zero in the cold region.
Notice that, according to Eqs. (\ref{actinh}), the constants $c_1$ and
$c_3$ are positive.

Obviously, $\et_s$ should be equal to the value of $\et$ at which
$v(\et) = 0$ in order for the pattern to be stationary. The value of
$\et_s$ must therefore satisfy \cite{ko:book,ko:ufn90,m:pre96}
\begin{equation} \label{max}
  \int_{\th_{s1}}^{\th_{s3}} q(\th, \et_s) d \th = 0.
\end{equation}
For small $\ett$ the function $v(\et)$ may be linearized around $\et_s$,
so we obtain \cite{m:pre96}
\begin{equation} \label{vl}
  v = {\ep B \ett \over \al Z a},
\end{equation}
where
\begin{equation} \label{B}
  B = -a \int_{\th_{s1}}^{\th_{s3}} q'_\et(\th, \et_s) d \th,
\end{equation}
and
\begin{equation} \label{Z}
  Z = \int_{\th_{s1}}^{\th_{s3}} \sqrt{ -2 U_\th} d \th, ~~~ U_\th = -
  \int_{\th_{s1}}^\th q(\th, \et_s) d \th.
\end{equation}
The constants $B$ and $Z$ are of order 1. Notice that, according to Eq.
(\ref{actinh}), the value of $B$ is positive.

Although Eqs. (\ref{v}) and (\ref{smlin}) with (\ref{vl}) are simpler
than the original interfacial dynamics equations, they are still
difficult to deal with. It appears, however, that these equations can be
further simplified for treating multidomain patterns by introducing some
averaged variables. We will outline this procedure in this section and
demonstrate its application to the periodic patterns in the subsequent
sections.

Let us introduce the number of domains per unit volume $n$ and the
radius of the domain $\rho$ (in one dimension $\rho$ is the half-width
of a domain). If we now average Eq. (\ref{smlin}) over the volume of
size $s$ such that $\rho \lesssim s \ll 1$, we will obtain the
``coarse-grained'' equation for the average value of the inhibitor $\etb
= \langle \ett \rangle$
\begin{equation}
  \label{coarse}
  {\partial \etb \over \partial t} = \Delta \etb - c_1 \etb ( 1 + c_2 n
  \rho^d \Omega_d ) - Q(\th_{s1}, \et_s) - a n \rho^d \Omega_d,
\end{equation}
where $\Omega_d$ is the volume of the $d$-dimensional unit sphere
($\Omega_1 = 2$), and
\begin{equation}
  \label{c2}
  c_2 = (c_3 - c_1)/c_1
\end{equation}
measures the asymmetry between the hot and the cold domains. According
to Eq. (\ref{actinh}), the value of $c_2 > -1$. Besides $\etb$, there is
a local contribution to $\ett$ due to the variation of $\ett$ on the
length scales of $\rho \ll 1$. Let us introduce
\begin{equation} \label{he}
  \hat{\et} = \ett - \etb
\end{equation}
Since $\hat{\et}$ varies on the short length scales, the terms
proportional to $\et$ in the right-hand side of Eq. (\ref{smlin}) are
small compared to the Laplacian as well.  Also, when the characteristic
time scale of variation of $\hat{\eta}$ is much greater than $\rho^2$,
as is the case in all interesting situations (see the following
sections), the time derivative in Eq.  (\ref{smlin}) is small compared
to the Laplacian. Subtracting Eq.  (\ref{coarse}) from Eq.
(\ref{smlin}) and neglecting all these terms, we obtain
\begin{equation}
  \label{hat}
  \Delta \hat{\et} = a \{ I(x) - \langle I(x) \rangle \}, ~~~\langle
  \hat{\et} \rangle = 0.
\end{equation}
From this equation one can obtain the value of $\hat{\et} = \hat{\et}_s$
in the wall of an individual domain, so from Eqs. (\ref{v}) and
(\ref{vl}) follows that the equation for the radius is
\begin{equation}
  \label{rad}
  {\partial \rho \over \partial t} = - {\ep^2 (d - 1) \over \al \rho} +
  {\ep B ( \etb + \hat{\et}_s) \over \al Z a}.
\end{equation}
The variables $\etb$, $\hat{\et}_s$, and $\rho$ may now be considered as
space- and time-dependent on the length scales much greater than $s$.
Thus, we have a closed set of equations for these coarse-grained
variables, so the number of the relevant dynamical variables in the
problem is considerably reduced.

\section{one-dimensional periodic strata of small period}

Let us consider the one-dimensional periodic strata of the period $\lp
\ll 1$ (Fig. \ref{strata}). In this case $n = \lp^{-1}$, so Eq.
(\ref{coarse}) becomes
\begin{equation} \label{eb}
  \ptl{\etb}{t} = \ptl{^2 \etb}{x^2} - c_1 \etb (1 - 2 c_2 \lp^{-1} \rho
  ) - 2 a \lp^{-1} (\rho - \rho_0),
\end{equation}
where we introduced
\begin{equation} \label{rb0}
  \rho_0 = - \lp Q(\th_{s1}, \et_s)/2 a.
\end{equation}
Equation (\ref{hat}) for a one-dimensional pattern with the period $\lp$
becomes
\begin{equation} \label{hee}
  {d^2 \hat{\et} \over dx^2} = a I(x) - 2 a \lp^{-1} \rho, ~~~\langle
  \hat{\et} \rangle = 0
\end{equation}
with neutral boundary conditions at $x = \pm \lp/2$ (because of the
translational invariance, we may choose the center of the domain to be
at $x = 0$). A straightforward calculation gives
\begin{equation} \label{hes}
  \hat{\et}_s = a \left(- \frac{\rho \lp}{6} + \rho^2 - \frac{4
    \rho^3}{3 \lp} \right).
\end{equation} 
Rescaling the variables $\etb$ and $\rho$
\begin{equation} \label{scale}
  \etb' = \etb / a \lp^2, ~~~ \rb = \rho / \lp,
\end{equation}
and introducing the quantities
\begin{equation} \label{new}
  \tau_1 = \al Z / \ep B \lp, ~~ \omega_0 = \sqrt{2 \ep B / \al Z \lp}.
\end{equation} 
we write Eqs. (\ref{rad}) and (\ref{eb}) (dropping the primes) as
\begin{equation} \label{ets}
  \ptl{\etb}{t} = \ptl{^2 \etb}{x^2} - c_1 \etb (1 + 2 c_2 \rb) -
  \omega_0^2 \tau_1 (\rb - \rb_0),
\end{equation}
\begin{equation} \label{rbs} 
  \tau_1 \ptl{\rb}{t} = \etb - \frac{\rb}{6} + \rb^2 - \frac{4
    \rb^3}{3}.
\end{equation}
The parameters of the original reaction-diffusion system [Eqs.
(\ref{act}) and (\ref{inh})] enter these equations only through
constants $\om_0$, $\tau_1$, and $c_2$ (the value of $c_1$ can be
absorbed in the definitions of $\tau_1$ and $\om_0$), and the excitation
level enters mainly through $\rb_0$. Note that since, according to its
definition, the value of $2 \rho$ cannot exceed $\lp$, we have $0 < \rb
< \frac{1}{2}$. Also note that Eqs.  (\ref{ets}) and (\ref{rbs}) with
the spatial derivative equal to zero describe the oscillations of a
single domain in the system of size $\lp \ll 1$.

Equations of the type of Eqs. (\ref{ets}) and (\ref{rbs}) have been
extensively studied in the context of oscillatory chemical reactions
\cite{cross93,field,mikhailov,kapral,murray,kuramoto}. It was shown that
these systems host a great richness of dynamic patterns, such as
self-sustained uniform oscillations, traveling waves, target patterns
and spirals (in higher dimensions), and spatio-temporal chaos
(turbulence).  This is also true for Eqs. (\ref{ets}) and (\ref{rbs}).
However, one should keep in mind that these equations describe the
dynamics of the already existing multidomain pattern, in other words,
the dynamical patterns described by Eqs. (\ref{ets}) and (\ref{rbs})
will be seen ``on top'' of the stationary multidomain (Turing) patterns.
This is an important distinction from the oscillatory systems, such as
oscillatory chemical reactions, in which the dynamic patterns appear on
top of the stationary homogeneous state.

We are interested in the collective oscillations of the domains in the
pattern. According to Eqs. (\ref{ets}) and (\ref{rbs}), the
characteristic frequency of these oscillations is equal to $\om_0$.
In view of Eq.  (\ref{new}), the period of oscillations is the
shortest time scale in the problem if
\begin{equation}
  \label{al}
  \ep \lp^3 \ll \al \ll \ep \lp^{-1}.
\end{equation}
Recall that in deriving Eq. (\ref{hat}) we neglected the time derivative
of $\hat{\et}$. This is justified if $\om_0 \ll \lp^{-2}$, what is
equivalent to $\al \gg \ep \lp^3$, so the first condition in Eq.
(\ref{al}) is in fact a necessary condition for Eqs. (\ref{ets}) and
(\ref{rbs}) to be valid.

Equations (\ref{ets}) and (\ref{rbs}) have a trivial solution $\rb =
{\rm const}$, $\etb = {\rm const}$ which corresponds to the stationary
multidomain pattern (strata). For $\al$ satisfying the condition in Eq.
(\ref{al}) we have $\om_0 \gg 1$, so the value of $\rb$ for the
stationary pattern is close to $\rb_0$. Linear stability analysis of
Eqs. (\ref{ets}) and (\ref{rbs}) shows that at $\tau_1 > \tau_{1c}$,
where
\begin{equation} \label{tau1c}
  \tau_{1c} \cong {12 \rb_0 - 24\rb_0^2 - 1 \over 6 c},
\end{equation}
and
\begin{equation}
  \label{c}
  c = c_1 (1 + 2 c_2 \rb_0) > 0,
\end{equation}
this solution is stable. At $\tau_1 = \tau_{1c}$ it looses stability
with respect to the uniform oscillations with the frequency $\om \cong
\om_0$ and the wave vector $k = 0$. The uniformly oscillating solution
corresponds to the pattern in which all domains are oscillating in phase
with the frequency $\om_0$. This is a supercritical Hopf bifurcation
(see below). In view of Eq. (\ref{new}) and (\ref{tau1c}), the value of
$\tau_{1c} \sim 1$, so in terms of the original variables we have $\al_c
\sim \ep \lp$ and $\om_0 \sim \lp^{-1}$.  Also, according to Eq.
(\ref{tau1c}), the value of $\tau_{1c} > 0$ only if
\begin{equation}
  \label{rmm}
  \frac{1}{4} \left( 1 - \frac{1}{\sqrt{3}} \right) < \rb_0 <
  \frac{1}{4} \left( 1 + \frac{1}{\sqrt{3}} \right),
\end{equation}
i.e., $0.11 < \rb_0 < 0.39$. This is a completely universal result for
all one-dimensional strata of small period in K$\Omega$N systems. It is
easy to see from Eq. (\ref{tau1c}) that if the condition of Eq.
(\ref{rmm}) is satisfied, for any acceptable value of $c_2$ there exists
$\tau_{1c} = \tau_{1c}^{max}$, such that the stationary pattern is
stable for $\tau_1 > \tau_{1c}^{max}$ for any value of $\rb_0$. If the
condition in Eq.  (\ref{rmm}) is not satisfied, the pattern will still
become unstable when $\al \sim \ep^2$
\cite{ko:book,ko:ufn89,ko:ufn90,mo1:pre96}. This will happen at the
values of $\al$ which are much smaller than in the former case.  Indeed,
according to Eq.  (\ref{tau1c}), the oscillatory instability occurs at
$\al \gg \ep^2$ since $\lp$ is always much greater than $\ep$
\cite{ko:book,ko:ufn90}.

To study the destabilization in more detail we will use the fact that
the period of the oscillations $\om_0^{-1}$ is much smaller than the
characteristic relaxation time $\tau_1$, so one can apply the method of
Bogoliubov and Mitropolsky to Eqs.  (\ref{ets}) and (\ref{rbs})
\cite{bm}. To do this, we will use Eq.  (\ref{rbs}) to eliminate $\etb$
from Eq. (\ref{ets}) and write
\begin{equation}
  \label{anz}
  \rb = \rb_0 + R \cos( \om_0 t + \Theta),
\end{equation}
where $R$ and $\Theta$ are slowly varying functions of $t$ and $x$.
Substituting Eq. (\ref{anz}) into Eq. (\ref{ets}), multiplying it by
$\sin( \om_0 t + \Theta)$ and $\cos( \om_0 t + \Theta)$ and integrating
over the period, we will get, respectively,
\begin{equation} \label{ampl}
  { \partial R \over \partial t} = {c R \over 2 \tau_1} (\tau_{1c} -
  \tau_1) - {R^3 \over 2 \tau_1} + \frac{1}{2} {\partial^2 R \over
    \partial x^2} - \frac{R}{2} \left( \frac{\partial \Theta}{\partial
    x} \right)^2,
\end{equation}
and
\begin{equation}
  \label{phase}
  {\partial \Theta \over \partial t} = \frac{1}{2} {\partial^2 \Theta
    \over \partial x^2} + \frac{1}{R} {\partial R \over \partial x}
  {\partial \Theta \over \partial x}.
\end{equation}
In deriving Eqs. (\ref{ampl}) and (\ref{phase}) we kept only the leading
terms.

Equations. (\ref{ampl}) and (\ref{phase}) are equivalent to the
equations for the amplitude and phase for the well-known
Newell-Whitehead equation \cite{newell69}. An important class of
solutions of this equation is $R = R_0$ and $\Theta = k x$, where
\begin{equation} \label{r0}
  R_0 = \sqrt{c (\tau_{1c} - \tau_1) - \tau_1 k^2 }
\end{equation}
and $k < \sqrt{c (\tau_{1c} - \tau_1) / \tau_1}$. It represent the
nonlinear dispersionless traveling waves of the phase of the domain
oscillations with the wave vector $k$ which are stable if $k < \sqrt{c
  (\tau_{1c} - \tau_1) / 3 \tau_1}$. A particular solution with $k = 0$
corresponds to the synchronous oscillations of the domains. In fact, if
the boundary conditions for the original reaction-diffusion system
(\ref{act}) and (\ref{inh}) are neutral, as in most of the real
situations, this is the only admissible plane wave solution. Since it is
stable, a variety of initial conditions will evolve into it, causing the
oscillations of different domains to synchronize. This means that the
synchronization of oscillations of different domains is the major
scenario of the development of the small-period patterns in one
dimension. Equations (\ref{ampl}) and (\ref{phase}) also admit a
solution in the form of a domain boundary (kink), upon going through
which the phase $\Theta$ changes by $\pi$.  These are the only stable
stationary solutions of Eqs.  (\ref{ampl}) and (\ref{phase}), so any
initial condition will in general produce a collection of regions in
which the domains oscillate in phase. In the regions next to each other
the domains will oscillate in antiphase.  At $\tau_1 < \tau_{1c}$ Eqs.
(\ref{ampl}) and (\ref{phase}) also have the solutions in the form of
the wave in which the stable state $R = R_0$ with $k = 0$ and $\Theta =
const$ invades the unstable state $R = 0$. This wave corresponds to the
onset of the synchronous oscillations in the unstable stationary strata
from a localized initial condition. The propagation velocity of this
wave is $v^* = \sqrt{c (\tau_{1c} - \tau_1 ) / \tau_1}$
\cite{vansaarloos88}.

It is clear that small local inhomogeneities, both intrinsic and
extrinsic, may play the role of the organizing centers for the
one-dimensional analog of the target patterns: waves traveling away from
the inhomogeneity in both directions. An example of an intrinsic
inhomogeneity here is a non-uniform distribution of the domains along
the $x$-axis. In this case instead of the constant density of domains we
have a smooth function $n(x)$, which can also vary in time and can be
considered as an extra degree of freedom in the problem. The density
must satisfy the continuity equations
\begin{equation}
  \label{n}
  \ptl{n}{t} + \ptl{}{x}(n {\rm v}) = 0,
\end{equation}
where
\begin{equation}
  \label{vv}
  {\rm v} = \frac{\rb}{\tau_2} \ptl{\etb}{x}
\end{equation}
is the velocity of the domain as a whole due to the gradient of $\etb$
[see Eq. (\ref{vl})], and
\begin{equation}
  \label{tau2}
  \tau_2 = \al Z / \ep B \lp^3
\end{equation}
is the characteristic time scale of the density variation. For $\tau_1
\sim 1$ we have $\tau_2 \sim \lp^{-2}$, so the density relaxes on much
longer time scale and any non-uniform distribution of $n$ is frozen on
the time scale of the relaxation of the amplitude and phase. This means
that by setting a non-uniform density of domains as the initial
condition one can produce complex spatio-temporal patterns of the domain
oscillations which will not synchronize for a very long time.

From the definition of $R$ follows that it lies in the interval from
$\rb_0$ to $\frac{1}{2} - \rb_0$. If $R$ exceeds one of these values,
the domains will either merge or collapse in the course of the
oscillations. On the other hand, according to Eq. (\ref{r0}), the value
of $R_0$ may exceed either $\rb_0$ or $\frac{1}{2} - \rb_0$ for some
values of $\tau_1$. This can be seen from Fig.  \ref{diag1d}, in which
the bottom solid line corresponds to the values of $\tau_1$ at which
this happen. Figure \ref{diag1d} thus shows that the synchronization of
the domain oscillations occurs only in the relatively narrow region of
the system's parameters. The merging or collapse of the domains will
result in the breakdown of the collective domain oscillations and
collapse of the strata. Indeed, suppose that at some moment every two
domains merged into one. This may be viewed as doubling of the pattern's
period $\lp$.  According to Eq.  (\ref{new}), this will result in the
decrease of $\tau_1$ and further increase of $R_0$.  Therefore, the
process of domain merging will have an avalanche character and lead to a
complete destruction of the strata. The way the breakdown of the strata
will follow will depend on whether or not the homogeneous state of the
systems is stable with respect to the uniform self-oscillations of the
homogeneous state at this value of $A$. If the homogeneous state of the
system is unstable, the breakdown will result in the onset of the
uniform {\em relaxation} self-oscillations (the latter follows from the
fact that since $\tau_{1c} \lesssim 1$ and $\ep \ll 1$ we have $\al \ll
1$ and, therefore, relaxation oscillations
\cite{ko:book,ko:ufn89,ko:ufn90}). If the homogeneous state of the
system is stable, the breakdown will result in the formation of
traveling domain patterns (AS) or in the collapse of the strata into the
homogeneous state.

The scenarios of the evolution of one-dimensional strata of small period
discussed in the previous paragraph are precisely what we see in the
numerical simulations of a concrete model. We also find a detailed
quantitative agreement between the limit cycle oscillations of a single
domain determined by Eqs. (\ref{ets}) and (\ref{rbs}) and the results of
the numerical simulations of a reaction-diffusion model for small enough
$\ep$ and $\lp$. 

Before concluding this section, let us discuss another dynamic behavior
of strata in one dimension. As was recently shown by Osipov, for $\al
\lesssim \ep$ periodic patterns may undergo the instability leading to
the formation of the traveling strata (a wave train) \cite{o:pd96}.
Osipov obtained the general criterion for this instability. In the case
of one-dimensional strata of small period it is possible to calculate
the threshold value of $\al$, since the spatial variation of $\et$ is
given by $\hat{\et}(x)$. Substituting it into the criterion (Eq. (4.5)
of Ref.  \cite{o:pd96}), we obtain that the transformation of the
stationary strata to traveling occurs at $\al_T \sim \ep \lp^3$ One can
see that this instability occurs for much smaller values of $\al$ than
the oscillatory instability for which $\al_c \sim \ep \lp$.  Therefore,
the instability leading to the formation of the traveling pattern can
never be reached for the one-dimensional strata of small period.

Throughout the analysis presented above we used the fact that the period
of the pattern is small compared to the length of the variation of the
inhibitor and used $\lp$ as the small parameter in the expansions. The
smallness of the parameter $\lp$ greatly simplified the treatment of the
domain interactions and lead to the universal results which are
practically independent of the concrete nonlinearities of the system.
This is a general property of the KN and K$\Omega$N systems
\cite{mo1:pre96,mo2:pre96,m:pre96}. Unfortunately, no such treatment is
possible for the one-dimensional periodic patterns whose period is of
order 1. In the latter case one is faced with the problem of treating
the dynamics of each individual domain separately and taking into
account the complex memory effects associated with the domain
interactions, which cannot be solved in general. So, the problem of the
interaction and dynamics of the one-dimensional patterns consisting of
the domains whose size if of order 1 remains open. This problem,
however, does not exist in higher dimensions, if $\ep$ is small enough.
In this case the domain sizes are necessarily small since the domains
whose size is greater than $\ep^{1/3}$ are always unstable with respect
to the transverse instability of their walls and split into smaller
domains \cite{mo1:pre96,m:pre96}.

\section{higher-dimensional multidomain patterns}

Let us now study periodic multidomain patterns in higher dimensions. For
definiteness we will consider a two-dimensional hexagonal pattern of
disk-shaped domains, the most frequent Turing pattern in chemical
experiments \cite{kapral}. As was shown by Muratov and Osipov, in two
dimensions a stable stationary multidomain pattern must consist of the
domains whose size is of order $\ep^{1/3}$ \cite{mo1:pre96,mo2:pre96}.
Since the period of the pattern is of the same order as the domain size,
it can be used as a natural small parameter, so all the ideas of the
analysis of the previous section should apply to the two-dimensional
multidomain patterns.

Let us proceed with the derivation of the equations describing the
collective oscillations of the domains in the pattern. For a hexagonal
pattern with the period $\lp$ the density of the domains is $n = 2 /
\sqrt{3} \lp^2$, so Eq. (\ref{coarse}) becomes
\begin{equation}
  \label{eb2}
  \ptl{\etb}{t} = \Delta \etb - c_1 \etb \left(1 + {2 \pi c_2 \rho^2
    \over \sqrt{3} \lp^2} \right) - {2 \pi a (\rho^2 - \rho_0^2) \over
    \sqrt{3} \lp^2},
\end{equation}
where we introduced
\begin{equation}
  \label{r02}
  \rho_0^2 = - {\sqrt{3} \lp^2 Q(\th_{s1}, \et_s) \over 2 \pi a}.
\end{equation}
Because of the anisotropy of the hexagonal lattice the shape of the
domains will in general slightly deviate from the disk shape. It is
clear that this deviation is small and is smaller when the radius of the
domains is smaller. So, for the sake of simplicity and without
significantly affecting the results of the analysis, we will ignore the
effects of the anisotropy and consider the domains to be ideally
circular. Also, instead of considering the problem for $\hat{\et}$ in
the hexagonal cell, we will solve it for a circular domain of radius
$\lp / 2$. Thus, we have
\begin{equation}
  \label{hat2}
  \frac{d^2\hat{\et}}{d r^2} + \frac{1}{r} \frac{d \hat{\et}}{d r} = a
  I(r) - \frac{4 a \rho^2}{\lp^2},~~~\langle \hat{\et} \rangle = 0,
\end{equation}
where $r$ is the radial coordinate and the boundary conditions are
neutral at $r = 0$ and $r = \lp/2$. Solving this equation for
$\hat{\et}$ and calculating $\hat{\et}_s = \hat{\et}(\rho)$, we get
\begin{equation}
  \label{hats2}
  \hat{\et}_s = a \left( \frac{3 \rho^2}{8} - \frac{3 \rho^4}{2\lp^2} +
  \frac{\rho^2}{2} \ln \frac{2 \rho}{\lp} \right),
\end{equation}
so, Eq. (\ref{rad}) becomes
\begin{equation}
  \label{r2}
  \ptl{\rho}{t} = - \frac{\ep^2}{\al \rho} + \frac{\ep B}{\al Z a}
  \left( \etb + \frac{3 \rho^2}{8} - \frac{3 \rho^4}{2 \lp^2} +
    \frac{\rho^2}{2} \ln \frac{2 \rho}{\lp} \right).
\end{equation}
Rescaling the variables according to Eq. (\ref{scale}) and dropping the
primes, from Eqs. (\ref{eb2}) and (\ref{r2}) we get
\begin{equation}
  \label{ets2}
  \ptl{\etb}{t} = \Delta \etb - c_1 \etb \left( 1 + \frac{2 \pi
    c_2}{\sqrt{3}} \rb^2 \right) - \frac{\pi}{\sqrt{3}} \om_0^2 \tau_1
  (\rb^2 - \rb_0^2),
\end{equation}
\begin{equation}
  \label{rbs2}
  \tau_1 \ptl{\rb}{t} = \etb - \frac{\bar{\ep}}{\rb} + \frac{3 \rb^2}{8}
  - \frac{3 \rb^4}{2} + \frac{\rb^2}{2} \ln 2 \rb,
\end{equation}
where
\begin{equation}
  \label{ebar}
  \bar{\ep} = \ep Z / B \lp^3.
\end{equation}
As before, $0 < \rb_0 < \frac{1}{2}$. The parameter $\bar{\ep}$ does not
appear in the one-dimensional case. It characterizes the stability of
the domains with respect to the transverse perturbations. Since the
curvature radius of the stable domains must be of order $\ep^{1/3}$
\cite{mo1:pre96,mo2:pre96}, the value of $\bar{\ep}$ is of order 1 for
the domains whose size is comparable with $\lp \sim \ep^{1/3}$. Notice
that Eqs. (\ref{ets2}) and (\ref{rbs2})without the space dependence and
with the coefficient $\pi / \sqrt{3}$ replaced by 2 in Eq. (\ref{ets2})
will describe the oscillations of a single domain in a circular region
of radius $\lp \ll 1$. This situation was realized in the experiment by
Haim {\em et al.} on the ferrocyanide-iodate-sulfite (FIS) reaction
\cite{haim96}. 

When the condition of Eq. (\ref{al}) is satisfied, Eqs. (\ref{ets2}) and
(\ref{rbs2}) have two time scales, as in the one-dimensional case: the
short time scale $\om_0^{-1}$ for the oscillations and the long time
scale $\tau_1$ for the relaxation of their amplitude and phase. However,
in contrast to the one-dimensional case, the analysis is complicated by
the fact that now the oscillations themselves are nonlinear. To proceed
further, we will use Eq. (\ref{rbs2}) to express $\etb$ in terms of
$\rb$ and substitute it into Eq. (\ref{ets2}), taking into account that
$\om_0 \gg 1$. We obtain
\begin{equation}
  \label{rr}
  \ptl{^2 \rb}{t^2} + {\pi \over \sqrt{3}} \om_0^2 (\rb^2 - \rb_0^2) =
  - \left( \frac{f(\rb)}{\tau_1} - \Delta \right) \ptl{\rb}{t},
\end{equation}
where
\begin{eqnarray}
  \label{f}
  f(\rb) = && c_1 \tau_1 \left( 1 + { 2 \pi c_2 \over \sqrt{3}} \rb^2
\right) \nonumber \\ && - {\bar{\ep} \over \rb^2} - \frac{5 \rb}{4} + 6
\rb^3 - \rb \ln 2 \rb.
\end{eqnarray}

The left-hand side of Eq. (\ref{rr}) is an equation of motion for a
particle of unit mass in the potential $V = {\pi \om_0^2 \over \sqrt{3}
  } \left( \frac{\rb^3}{3} - \rb_0^2 \rb \right)$. For $0 < \rb <
\frac{1}{2}$ it describes finite motion between $\rb = \rb_{min}$ and
$\rb = \rb_{max}$ with the characteristic frequency $\om_0$. The
right-hand side of Eq. (\ref{rr}) is a weak nonlinear friction force
which also contains the diffusion term that couples the oscillators in a
spatially distributed system.

As in the case of the one-dimensional patterns, the equilibrium solution
$\rb = \rb_0$ of Eq. (\ref{rr}) corresponds to the stationary
multidomain pattern. Equation (\ref{rr}) shows that the stationary
multidomain pattern becomes unstable (the friction $f$ becomes negative)
with respect to the fluctuation with $\om \cong \om_0$ and $k = 0$ when
$\tau_1 < \tau_{1c}$, where
\begin{equation}
  \label{tau1c2}
  \tau_{1c} = {4 \bar{\ep} + 5 \rb_0^3 - 24 \rb_0^5 + 4 \rb_0^3 \ln 2
    \rb_0 \over 4 c_1 \rb_0^2 \left( 1 + {2 \pi c_2 \over \sqrt{3}}
    \rb_0^2 \right) }.
\end{equation}
Going back to the original variables, we see that in the case of the
two-dimensional multidomain pattern the destabilization occurs at
\begin{equation}
  \label{al2}
  \al_{c} \sim \ep^{4/3}.
\end{equation}
Note that the value of $\al_c$ is smaller than $\al_\om \sim \ep$ at
which a localized domain (AS) is destabilized with respect to pulsations
\cite{ko:book,ko:ufn89,mo1:pre96}.

Figure \ref{tau2d} shows the possible forms of the dependences
$\tau_{1c}(\rb_0)$ obtained from Eq. (\ref{tau1c2}) for different values
of $\eb$. One can see from Eq. (\ref{tau1c2}) that for $\eb > 0.031$ for
any value of $\rb_0$ there exists a value of $\tau_1$ at which the
instability occurs.  When $0.00045 < \eb < 0.031$ the instability is
realized only for $\rb_0 < \rb_{a3}$ with $\rb_{a3} < \frac{1}{2}$,
whereas for $\eb < 0.00045$ the instability is realized for $\rb_0 <
\rb_{a1}$ and $\rb_{a2} < \rb_0 < \rb_{a3}$ (see Fig. \ref{tau2d}). When
$\eb > \eb^*$ (for example, $\eb^* = 0.0034$ for $c_2 = 0$) for a given
$\tau_1$ the instability is realized for a single value of $\rb_0$,
whereas for $\eb < \eb^*$ there are values of $\tau_1$ for which the
instability is realized for three values of $\rb_0$.

Let us take a closer look at Eq. (\ref{rr}). The behavior of the
oscillations is determined by the sign of the friction term $f$ in Eq.
(\ref{rr}). Three possible forms of this term as a function of $\rb$ are
shown in Fig. \ref{fric}. The oscillation amplitude decreases if $\rb$
remains in the domain where $f > 0$, or increases if $\rb$ is in the
domain where $f < 0$ (indicated by arrows in Fig.  \ref{fric}).  In the
situation of Fig. \ref{fric}(c) all values of $\rb$ correspond to the
amplification region, so the amplitude of any oscillations will increase
until $\rb_{max}$ becomes equal to $\frac{1}{2}$ or until $\rb_{min} =
0$. In the first case the neighboring domains will merge while in the
second the domain will collapse. Both these effects will cause the
restructuring of the pattern and significant changes in the collective
domain oscillations.  In the situation of Fig. \ref{fric}(b) the
oscillations are amplified when $\rb < \rb_{c3}$ and attenuated when
$\rb > \rb_{c3}$. The stationary multidomain pattern will be stable when
$\rb_0 > \rb_{c3}$ or unstable when $\rb_0 < \rb_{c3}$. Depending on the
signs of the first and the second derivatives of $f(\rb)$ the
bifurcation at $\rb_0 = \rb_{c3}$ may be both supercritical and
subcritical (this is a Hopf bifurcation). It is clear, however, that
deeper into the unstable regime the oscillations will be amplified until
they collapse when $\rb_{min} = 0$ even in the case of the supercritical
bifurcation.

The most interesting situation is shown in Fig. \ref{fric}(a). There the
oscillations are amplified when $\rb < \rb_{c1}$ or $\rb_{c2} < \rb <
\rb_{c3}$ and attenuated for the rest of the values of $\rb$, so the
pattern is stable only for $\rb_{c1} < \rb_0 < \rb_{c2}$ or $\rb_0 >
\rb_{c3}$. The bifurcation at $\rb_{c3}$ is supercritical, while the
bifurcation at $\rb_0 = \rb_{c1}$ is subcritical. This can be easily
seen from the following argument. By varying the value of $\tau_1$ one
can make the minimum of $f$ between $\rb_{c2}$ and $\rb_{c3}$
arbitrarily shallow, thus controlling the amplification of the
oscillations in the region $\rb_{c2} < \rb < \rb_{c3}$. Since in this
case the small region of amplification is surrounded by a finite region
of dissipation, the bifurcation at $\rb_{c2}$ and $\rb_{c3}$ will be
supercritical. Since the first and the second derivatives of $f$ at
$\rb_{c1}$ are opposite to the one at $\rb_{c2}$, the bifurcation at
$\rb_{c1}$ is subcritical. If the value of $\tau_1$ is decreased, the
minimum of $f$ between $\rb_{c2}$ and $\rb_{c3}$ gets deeper, and at
some value of $\tau_1$ the bifurcation at $\rb_{c2}$ becomes
subcritical. Upon further decreasing the value of $\tau_1$ the limit
cycle associated with the subcritical bifurcation at $\rb_{c2}$ will
become unstable and lead to the collapse of the oscillations. Thus, the
most stable limit cycle oscillations should be expected when $\rb_{c2} <
\rb_0 < \rb_{c3}$ and the minimum of $f$ between $\rb_{c2}$ and
$\rb_{c3}$ is sufficiently shallow. The analysis of the bifurcation
types for different values of $\eb$ and $\rb_0$ is summarized in Fig.
\ref{reg2} (for simplicity we used $c_2 = 0$).

Up to now we discussed the oscillations of a single domain in the
multidomain pattern. What we found is that in certain situations the
synchronous oscillations of all domains in the pattern can be stable.
This naturally implies that the synchronously pulsating multidomain
pattern is an attractor and the synchronization of the domain
oscillations is therefore one of the major scenarios of the development
of the stationary multidomain patterns. However, the situation is
reacher in two dimensions because of the possibility of other classes of
stable solutions, such as target patterns and spiral waves. In
principle, it is possible to use the fact that $\om_0^{-1} \ll \tau_1$
and obtain the equations for the amplitude and phase by averaging over
the trajectories of the frictionless oscillators in this case as well.
This calculation however, cannot be carried out in the analytical form.
On the other hand, in a wide region of the parameters the bifurcation of
the stationary pattern is supercritical, so in its vicinity the dynamics
of the pattern is described by the complex Landau-Ginzburg equation. In
the appropriately scaled time and space variables this equation in the
considered case has the form
\begin{equation}
  \label{clg}
  \ptl{W}{t} = \Delta W + W - (1 + i b) |W|^2 W,
\end{equation}
where $W$ is appropriately normalized complex amplitude and
\begin{equation}
  \label{b}
  b = {4 \cdot 2^{1/2} \cdot 3^{3/4} \pi^{1/2} \rb_0^{5/2} \om_0
    \tau_{1c} \over {16 \pi \over \sqrt{3} } c_1 c_2 \tau_{1c} \rb_0^4 -
    56 \eb + 216 \rb_0^5 + \rb_0^3 + 4 \rb_0^3 \ln 2 \rb_0 }.
\end{equation}

The complex Landau-Ginzburg equation has been the subject of intensive
studies for the last two decades (see, for example, Refs.
\cite{cross93,kuramoto,vansaarloos92,chate94}). Specifically, for the
case of Eq. (\ref{clg}) Hagan obtained the solution in the form of a
steadily rotating spiral wave \cite{hagan82}. In our case this spiral
wave will be seen on top of the stationary multidomain pattern.
Precisely this phenomenon was observed by Boissande, Dulos, and De
Kepper in the experiments with the Turing patterns in the
chlorite-iodide-malonic acid (CIMA) reaction \cite{boissande}.

Kuramoto and Koga showed that the spiral waves in Eq. (\ref{clg}) become
unstable at sufficiently large $b$ \cite{kuramoto81}. They showed that
the spiral breakup leads to the formation of a chaotic spatio-temporal
pattern --- spiral turbulence. This will be the case in the situation we
study.  Indeed, throughout the analysis we assumed that $\om_0 \tau_1 \gg
1$, what means that $b \gg 1$, so in general the spiral wave solution is
unstable. Note, however, that the value of $b$ decreases as $\lp$
increases, so when $\lp$ (and $\ep$) is not very small, the value of $b$
may become small enough, so that the spiral wave solution is stable.
This can also be achieved by making the coupling constant $B$ smaller.

Since generally $b \gg 1$, plane wave solutions will be stable only in a
narrow range of the wave vectors around $k = 0$ \cite{kuramoto}. This is
because, in contrast to the one-dimensional case, the dispersion of the
plane waves is high. However, for Eq. (\ref{clg}) the Benjamin-Feir
instability is not realized, so the synchronous oscillations of the
domains are always stable in the spatially distributed system close to
the bifurcation point.

The stationary multidomain patterns are never perfect. The defects of
the pattern will work as initiators for the spiral waves, which will in
turn break up and produce more spiral waves invading the entire pattern,
so the formation of the chaotically oscillating multidomain pattern is
the more likely scenario for the development of the unstable stationary
multidomain pattern in two dimensions. 

The imperfections of the multidomain pattern may also work as the
guiding centers generating target waves of the oscillation phase
\cite{kuramoto}. An argument similar to the one applied to the
one-dimensional strata may be applied here. The target waves may be
initiated by the smooth spatial variations of the density of domains.
The equation for the density will be given by Eqs. (\ref{n}) and
(\ref{vv}), if one replaces $\partial / \partial x$ by $\nabla$, so the
time scale of the density variation will once again be much longer than
$\tau_1$, and, therefore, the density can be considered frozen on the
latter time scale. However, since the dispersion of the waves in the
considered situation is high, the target patterns are likely to be
unstable, what should also lead to the stochastization of the domain
oscillations.

It is clear that qualitatively the scenarios discussed above are also
realized when $\tau_1$ is not very close to $\tau_{1c}$. Of course, for
arbitrary $\tau_1$ one should also take into account the possibility of
the collapse or merging of the domains in the course of the
oscillations. To determine the values of $\tau_1$ at which the collapse
or merging of the domains occur for a given value of $\rb_0$ one can
consider the oscillator in Eq. (\ref{rr}) without the space dependence.
The collapse or merging of the domains occurs when $\rb_{min} = 0$ or
$\rb_{max} = \frac{1}{2}$, respectively. The values of $\rb_{min}$ and
$\rb_{max}$ can be obtained from the condition of energy balance for the
steady oscillations. Recalling that the friction is weak, from the
elementary mechanics we get
\begin{equation}
  \label{bal2}
  \int_{\rb_{min}}^{\rb_{max}} f(\rb) \sqrt{2 \left( E - {\pi \om_0^2
      \over \sqrt{3}} \left( \frac{\rb^3}{3} - \rb_0^2 \rb \right)
  \right) } = 0,
\end{equation}
where $E$ is the energy of the oscillator corresponding to the values of
$\rb_{min}$ and $\rb_{max}$. We solved this equation numerically in the
case $c_2 = 0$. Figure \ref{fampl} shows a typical dependence of the
amplitude of the oscillations on $\tau_1$ for a given value of $\eb$ in
the case of the supercritical bifurcation ($\rb_{min}$ and $\rb_{max}$
correspond to the left and the right portions of the solid curve,
respectively). From this figure it can be seen that the limit cycle
oscillations are stable only if $\tau_{1d} < \tau_1 < \tau_{1c}$. If
$\tau_1 < \tau_{1d}$, the amplitude of the oscillations will increase
until the domain collapses or merges with the neighbors. Several
possibilities exist for the values of $\rb_{min}^d$ and $\rb_{max}^d$.
The value of $\tau_{1d}$ may happen to be negative, in this case the
stability of the limit cycle oscillations will depend on whether the
value of $\rb_{max}$ becomes greater than $\frac{1}{2}$ at some value of
$\tau_1$. The oscillations will be stable for $\tau_1$ greater than this
value if this is the case, or stable for all values of $\tau_1$ in the
opposite case. The value of $\rb_{max}^d$ may also happen to be greater
than $\frac{1}{2}$. In this case the region of the stability of the
domain oscillations will also be narrower. Notice that $\rb_{min}^d$ is
always greater than zero because of the singular character of $f(\rb)$
at $\rb = 0$. 

In the case of subcritical bifurcation the limit cycle oscillations are
unstable for any values of $\tau_1$, so collective oscillations of the
domains in the multidomain pattern always break down in this case. The
most likely scenario here is that some of the domains will shrink and
collapse whereas some will grow, what can be effectively thought of as
an increase of the pattern's period. As a result, the value of $\eb$,
which mainly determines the type of the bifurcation, will get smaller,
so in the new pattern of greater period the bifurcation may happen to be
supercritical. This means that the pattern may naturally evolve into a
state in which the collective domain oscillations are stable from more
or less arbitrary initial conditions, provided that the initial
condition consists of the alternating hot and cold domains of size less
or of order $\ep^{1/3}$.  It is also possible that the pattern that
forms in this process is stable with respect to the domain oscillations.
Yet another possibility is that in this process some of the domains will
merge, what will result in the formations of an irregular pattern, which
also may exhibit chaotic dynamics. The merging process may also lead to
the destruction of the multidomain pattern. We will get back to these
points when we will discuss the results of the numerical simulations of
a concrete model. 

In the analysis presented in this section we did not specify precisely
the range of the values of $\eb$. This range is determined by the
stability of the stationary multidomain pattern with respect to the
activator repumping and the transverse distortions of the walls of the
domains. When $\eb$ is greater than some value for a given $\rb_0$, the
pattern becomes unstable with respect to the activator repumping effect
leading to the doubling of the pattern's period, whereas when $\eb$ is
smaller than some critical value at a given $\rb_0$, the domains
destabilize with respect to the transverse instability of their walls
leading to the radially-nonsymmetric distortions of the domains
\cite{ko:book,ko:ufn90,mo2:pre96,m}. Both these instabilities occur when
$\eb \sim 1$ in the limit $\ep \rightarrow 0$ regardless of the value of
$\al$. It is possible to show \cite{m} that for the stable stationary
multidomain pattern in two dimensions the value of $\eb \sim 0.002$, so
the scenarios discussed above are indeed realized. Also, according to
the Osipov's criterion \cite{o:pd96}, the transformation of a stationary
multidomain pattern into traveling will occur when $\al_T \sim \ep^2 \ll
\al_c$.

For the purpose of completeness let us quote the equations obtained in
the case of the three-dimensional multidomain pattern consisting of
spherical domains situated on a closed-packing lattice. The derivation
follows along the same lines as the derivation in two dimensions.
Introducing
\begin{equation}
  \label{r03}
  \rho_0^3 = - { 3 \lp^3 Q(\th_{s1}, \et_s) \over 4 \pi \sqrt{2} a},
\end{equation}
and rescaling $\etb$ and $\rho$ according to Eq. (\ref{scale}), we
obtain 
\begin{equation}
  \label{ets3}
  \ptl{\etb}{t} = \Delta \etb - c_1 \etb \left( 1 + { 4 \pi \sqrt{2}
    \over 3} c_2 \rb^3 \right) - {2 \pi \sqrt{2} \over 3} \om_0^2 \tau_1
  (\rb^3 - \rb_0^3),
\end{equation}
\begin{equation}
  \label{rbs3}
  \tau_1 \ptl{\rb}{t} = \etb - {2 \eb \over \rb} - {\rb^2 \over 3} + {6
    \rb^3 \over 5} - {32 \rb^5 \over 15}. 
\end{equation}
The properties of these equations are essentially the same as in two
dimensions. 

\section{entrainment of the oscillations in a disordered pattern}

The multidomain patterns that we studied so far consisted of domains
situated on a regular lattice. We found that the motion of the domains
occurs on two time scales: the short time scale $\om_0^{-1}$ which is
proportional to the period of the oscillations of a single domain and
the long time scale $\tau_1$. The motion of the domains on the time
scale $\om_0^{-1}$ is conservative, so the time $\tau_1$ is associated
with the relaxation of the oscillator's energy. In two dimensions we
found a lot of the situations in which the collective domain
oscillations should become chaotic, either because of the nonlinear
dynamics of the oscillation amplitude and phase, or because of the
defects. This chaotic dynamics, however, is realized on the long time
scale and is essentially determined by the nonlinear relaxation
processes. Here we would like to ask the following question. Suppose
that instead of the domains of a fixed radius on a perfect lattice
(hexagonal in two dimensions), we have a random arrangement of the
domains with a distribution of radii. What will be the dynamics of
different domains on the short time scale $\om_0^{-1}$?

To proceed, we will consider a simplified situation and ignore the space
dependence of all dynamical quantities in the problem. One can think
that the domains are considered in the system whose size is much smaller
than 1.  Let $n(\rho, t)$ now be the distribution function of the domain
radii $\rho$. The distribution $n(\rho, t)$ must satisfy the Liuville
equation obtained from Eq. (\ref{rad}). Rescaling appropriately $n$,
$\etb$, $\rho$, and $t$, and neglecting the terms which are not
significant on the time scale $\om_0^{-1}$, from Eq. (\ref{coarse}) and
(\ref{rad}) we get
\begin{equation}
  \label{eee}
  \ptl{\etb}{t} = \rho_0^d - \int \rho^d n(\rho, t) d \rho,
\end{equation}
\begin{equation}
  \label{liuv}
  \ptl{n}{t} = \etb \ptl{n}{\rho},
\end{equation}
where $\rho_0^d$ is a constant which comes from the term $Q(\th_{s1},
\et_s)$ in Eq. (\ref{coarse}) and the distribution $n(\rho, t)$ is
normalized to unity:
\begin{equation}
  \label{norm}
  \int n(\rho, t) d \rho = 1.
\end{equation}
The last condition implies the conservation of the number of the domains
as a function of time. This condition is in fact not always satisfied,
since we have an absorbing boundary condition for $n$ at $\rho = 0$ and
even more sophisticated situation at large $\rho$ because of the
possibility of the domain merging. It is clear, however, that this
condition will be satisfied if the distribution $n$ has finite support
at all times and the possibility of merging is excluded. 

Let us introduce the quantities
\begin{equation}
  \label{mk}
  m_0 = 1, ~~~m_k(t) = \int \rho^k n(\rho, t) d \rho,
\end{equation}
where $k = 1, 2, \ldots, d$. Then from Eqs. (\ref{eee}) and (\ref{liuv})
follows:
\begin{equation}
  \label{mmm}
  \frac{d m_k}{d t} = - k \etb m_{k-1}, 
\end{equation}
\begin{equation}
  \label{mdd}
  \frac{d \etb}{d t} = \rho_0^d - m_d.
\end{equation}

Equations (\ref{mmm}) and (\ref{mdd}) can be solved exactly. One can use
Eq. (\ref{mmm}) with $k = 0$ to eliminate $\etb$ from Eq. (\ref{mmm})
with $k = 1$ and integrate this equation to get $m_2 = m_1^2 + C_1$,
where $C_1$ is a constant of integration. This equation, together with
Eq. (\ref{mmm}) with $k = 0$ can be used to integrate Eq.  (\ref{mmm})
with $k = 2$ to get $m_3 = m_1^3 + 3 C_1 m_1 + C_2$, and so forth. Thus,
the whole set of Eqs. (\ref{mmm}) and (\ref{mdd}) can be reduced to a
single equation for $m_1$. The constants of integration are determined
by the initial distribution function $n_0(\rho)$. If one knows the
solution of the equation for $m_1$, one then gets
\begin{equation}
  \label{nnn}
  n(\rho, t) = n_0 \bigr( \rho - m_1(t) + m_1(0) \bigl).  
\end{equation}

The first conclusion one draws from the solution of Eqs. (\ref{mmm}) and
(\ref{mdd}) is that the oscillations of different domains in a random
pattern are always entrained. This is in fact obvious right from the
start, since the rate of change of the radii for each domain is
proportional to $\etb$ which in turn depends integrally on the
distribution of radii. However, the dynamics of the domains is
determined by the initial distribution $n_0$ and may be qualitatively
different for different initial conditions, even if the parameters of
the system are the same. 

Let us see what happens when $d = 1, 2, 3$. In the case $d = 1$ the
equation for $m_1$ is
\begin{equation}
  \label{m1}
  {d^2 m_1 \over d t^2} + m_1 - \rho_0 = 0.
\end{equation}
One can see that for any distribution $n_0$ (with finite support and and
assuming that no domain merging occurs) the motion is that of a simple
harmonic oscillator. The energy of the oscillator is determined by the
value of $\etb$ at $t = 0$. One can see that dynamics of the
random pattern on the time scale $\om_0^{-1}$ is identical to the that
of the ideally periodic pattern.

In $d = 2$ and $d = 3$ the situation becomes somewhat more complicated.
In $d = 2$ we have
\begin{equation}
  \label{m2}
  {d^2 m_1 \over d t^2} + m_1^2 + C_1 - \rho_0^2 = 0.
\end{equation}
Here again the dynamics is equivalent to that of an ideal hexagonal
pattern, but the equilibrium point is shifted. The shift is determined
by $C_1 = \langle \rho^2 \rangle - \langle \rho \rangle^2$ at $t =
0$. One can see from Eq. (\ref{m2}) that when $C_1 > \rho_0^d$ the
oscillations are not possible. 
In $d = 3$ we have 
\begin{equation}
  \label{m3}
  {d^2 m_1 \over dt^2} + m_1^3 + 3 C_1 m_1 + C_2 - \rho_0^3 = 0,
\end{equation}
and the dynamics of the domains may be different from that of the
domains in an ordered pattern.

The main conclusion that follows from the analysis above is that the
disorder, especially small, should not significantly affect the dynamics
of the pattern and that the equations for $\etb$ and $\rb$ should
adequately describe the behavior of irregular multidomain patterns as
well. It is also clear that the distortions of the domain shapes will
not have significant effect either.

\section{simulations of a concrete model}

In this section we present the result of the numerical simulations of
the model with a simple cubic nonlinearity:
\begin{equation}
  q = \th^3 - \th - \et, \label{q} 
\end{equation}
\begin{equation}
  Q = \th + \et - A. \label{Q}
\end{equation}

Recently, Muratov and Osipov performed extensive numerical simulations
of this model in two dimensions \cite{mo2:pre96}. The emphasis of their
work was on the instabilities of the localized solitary structures (AS)
and the evolution of the patterns excited by a short localized external
stimulus in the stable homogeneous system. They were able to construct a
state diagram showing what kinds of patterns are realized for different
values of $\al$ and $A$ for a fixed value of $\ep \ll 1$ from localized
initial conditions. Here we perform numerical simulations of Eqs.
(\ref{act}) and (\ref{inh}) with (\ref{q}) and (\ref{Q}) with
non-localized initial condition.

The homogeneous state of the system under consideration is stable when
$|A| > 1/3 \sqrt{3} \simeq 0.19$. The various constants involved are
\cite{mo2:pre96,m:pre96}
\begin{equation}
  \label{const}
  a = 2, ~~B = 4, ~~Z = {2 \sqrt{2} \over 3}, ~~c_1 = \frac{3}{2}, ~~c_2
  = 0.
\end{equation}

Unfortunately, the direct simulations of Eqs. (\ref{act}) and
(\ref{inh}) with $\ep \ll 1$ in the system of sufficiently large size
are extremely time-consuming even on a very fast computer. In order for
the equations for $\rb$ and $\etb$ to be in quantitative agreement with
the simulations we should have $\ep \lesssim 0.01$. Nevertheless, it is
expected to have qualitative agreement with the predictions of the
preceding sections. All our simulations were performed with $\ep =
0.05$.

Our first observation is that if a stable stationary multidomain pattern
is taken as an initial condition in the run with sufficiently small
$\al$, the pattern will destabilize with respect to the synchronous
oscillations of the domains, and its evolution will depend on how small
the value of $\al$ is and on whether the homogeneous state of the system
is stable. When $\al$ is slightly below the threshold value and the
pattern's period is not very small, the oscillations of different
domains will synchronize. If the value of $\al$ is smaller, the domains
may start to merge during the oscillations, what will result in the
formation of an irregular pulsating pattern. If $\al$ is even smaller
and the homogeneous state of the system is stable, the pattern will
collapse into the homogeneous state after a few periods of oscillations
or transform into a turbulent pattern. The turbulence here is induced by
the self-replication of the domains and is qualitatively different from
the chaotic behaviors discussed above \cite{mo2:pre96}.  If the value of
$\al$ is small enough and the homogeneous state of the system is
unstable with respect to the uniform self-oscillations, the multidomain
pattern will collapse into the uniform self-oscillations.  This is in
agreement with the predictions of our analysis.

Besides the synchronous oscillations of the domains, we predicted
traveling waves of the oscillation phase. These traveling waves are
indeed realized in the simulations (Fig. \ref{wave}). One has to use the
special initial conditions to excite a traveling wave. The distributions
of $\th$ should consist of the domains whose radii vary smoothly along
the $x$-axis, and the distribution of $\et$ should follow this variation
with the phase difference of $\pi/2$.

In addition to the plane waves, target waves of the oscillation phase
are observed.  Figure \ref{chaos} shows a portion of a target wave
emanating from the top left corner of the system. The wave breaks down
after a while, as the domains begin to merge during the oscillations. As
a result, an irregular chaotically oscillating pattern forms in the
system.  Notice that eventually, when no merging occur any longer, the
average size of the domains becomes greater than the size of the domains
at the beginning. This can be interpreted as a self-organized increase
of the patterns period leading to the stabilization of the pattern's
oscillations.

In another simulation in which the value of $\al$ was smaller, the
target pattern broke down not only because of merging, but also because
of the collapse of smaller domains (Fig. \ref{chaos2}). As a result of
the domain collapse, patches of the almost homogeneous state formed in
the system. These patches are then invaded by the domains, what leads to
the formation of an irregular pulsating pattern. For smaller values of
$\al$ this may also lead to the formation of autowaves and turbulence
induced by self-replication of domains \cite{mo2:pre96}. 

In the simulations above we used a hexagonal pattern as the initial
condition. Let us now see what happens in the more realistic situation
when the multidomain pattern is disordered. Figure \ref{unst} shows the
evolution of such pattern. The initial stationary pattern was in fact
generated as a result of the Turing instability of the homogeneous state
at large $\al$. In the simulation of this figure the value of $\al$ is
small enough, so that the stationary multidomain pattern is unstable,
and a small disturbance was added to $\et$ in the vicinity of the top
left corner of the system in order to make the amplitude of the
oscillations in this region bigger. We see that at early times (the
first three plates in Fig. \ref{unst}) the oscillations of the domains
are indeed entrained, despite the differences in the domain sizes and
shapes. Notice that in this simulation the homogeneous state of the
system is unstable with respect to the uniform self-oscillations. One
can see that here the amplitude of the oscillations increases until
merging and collapse of the domains occur at the top left corner, which
are then followed by the formation of the uniform self-oscillations. The
uniform self-oscillations eventually invade the whole system and destroy
the multidomain pattern. Yet the uniform self-oscillations and the
multidomain pattern may coexist for quite a long time. This effect was
observed in the dynamics of the Turing pattern in the CIMA reaction
\cite{boissande}.

Above we considered the dynamics of the patterns consisting of circular
hot domains in the cold system. Clearly, qualitatively the same dynamics
is expected for the cold domains in the hot system. In fact, by changing
$\th \rightarrow -\th$, $\et \rightarrow -\et$, and $A \rightarrow -A$
in the concrete system under consideration we will transform hot domains
into cold and vice versa. A possibility exists for such oscillations of
the multidomain pattern in which the hot domains will transform into
cold domains and back. This process is illustrated in Fig. \ref{flip}.
There a stationary multidomain pattern which formed as a result of the
Turing instability at $A = -0.1$ when the hot domains are favorable was
taken as the initial condition. The simulation was performed at $A =
0.1$ when the cold domains are favorable. One can see the sequence of
transitions from the hot domains to cold and back during the
oscillations. A lot of the domains merge as their size increases, so the
multidomain pattern quickly transforms into an irregular pattern
consisting of many disconnected pieces. The connectivity of the domains
changes with time. After $t = 1.4$ the systems eneters a periodic cycle
in which the pattern changes periodically from hot to cold. In the
run with these parameters the pattern eventually collapsed into the
homogeneous self-oscillations. 

So, in summary, the collapse and merging of the domains appear to be the
major cause for the chaotization of the oscillations. The chaos here is
due to the underlying irregularity of the domains themselves. The
predicted chaotic behavior near the onset of synchronous domain
oscillations seems to be beyond the capabilities of the available
computational power. While we were able to see steady non-uniform
oscillations of a hexagonal multidomain pattern of period $\lp = 2.3$ at
$\ep = 0.1, \al = 0.05$, and $A = -0.1$ in the system $40 \times 40$,
the system was yet too small to identify these oscillations with chaos.

\section{conclusion}

Thus, in this paper we studied the dynamics of the multidomain patterns
of small period in K$\Omega$N systems. In order to simplify the problem
we took the advantage of the smallness of $\ep$, the natural small
parameter in the considered systems
\cite{ko:book,ko:ufn89,ko:ufn90,mo1:pre96}. In the limit $\ep
\rightarrow 0$ the dynamics of the pattern reduces to the free boundary
problem \cite{ohta89,m:pre96}. This problem may be further simplified by
using the smallness of the pattern's period. This approach is somewhat
limited in one dimension, but is always applicable to higher-dimensional
multidomain patterns, if $\ep$ is small enough. In the case of the
periodic patterns we managed to reduce the number of dynamical variables
involved to the average domain radius $\rho$ and the average value of
$\et$. The equations for these quantities turned out to be universal.
The nonlinearities of the original reaction-diffusion system enter only
via a few numerical constants of order 1. In $d \geq 2$ the dynamics
studied by us is in fact the asymptotic limit of the true dynamics as
$\ep \rightarrow 0$. Note, however, that the value of $\ep$ has to be
sufficiently small in order for this approach to be in quantitative
agreement with the actual dynamics. For example, for the concrete model
studied in Sec. VI this is the case only when $\ep \lesssim 0.01$ (see
also Ref. \cite{mo2:pre96}).  Nevertheless, the qualitative agreement is
good for $0.01 \lesssim \ep \ll 1$.

The analysis performed by us predicts both synchronization and chaos in
the collective domain oscillations of the multidomain patterns. There
are two types of chaos that are realized: the chaos associated with the
effects of the interplay of the dissipation and dispersion of the
nonlinear waves of the phase of the domain oscillations and the chaos
associated with the irregularity of the underlying multidomain pattern.
The first kind of chaos is shown to correspond to the intermittency and
defect turbulence in the complex Landau-Ginzburg equation. The second
kind of chaos is due to the breakdown of the averaged dynamics
description and is associated with the collapse and merging of the
domains and the destruction of the regular multidomain pattern.
Nevertheless, it is shown that even in an irregular pattern the domain
oscillations are synchronized locally, so the chaos is still realized on
the time and length scales of the dissipation processes.

Recall that $\ep = \sqrt{\al D_\th / D_\et}$, where $D_\th$ and $D_\et$
are the diffusion coefficients of the activator and the inhibitor,
respectively. According to its definition, the value of $\ep$ is small
as long as $\al$ is small and $D_\et \gtrsim D_\th$. However, if the
values of $D_\th$ and $D_\et$ are of the same order, as is the case in
typical experiments with the autocatalytic reactions \cite{epstein95},
this means that $\al \sim \ep^2$. At this relationship between $\al$ and
$\ep$ the stationary patterns are always unstable
\cite{ko:book,ko:ufn89,ko:ufn90,mo1:pre96,epstein95}, and only autowaves
will be realized. Our analysis also shows that this will be the case for
the multidomain patterns. In other words, one should have $D_\et \gg
D_\th$ in order for the stationary and more complex dynamic patterns
(not autowaves) to be feasible in an experiment. As was emphasized by
Epstein and Lengyel, in chemical systems, for example, one has to devise
special methods to make the ratio of the diffusion coefficients large
\cite{epstein95}. On the other hand, the value of $\al$ is determined by
the kinetics of the chemical reactions involved and can be easily made
small.

It is well known that a solitary pattern (AS) in one dimension would
destabilize and transform into a pulsating pattern when $\al \sim \ep$,
what implies that $D_\th / D_\et \sim \al$
\cite{ko:book,ko:ufn89,ko:ufn90,mo1:pre96,koga80,nishiura89}. The same
condition must be satisfied in order for the instability of the
homogeneous state of the system with respect to the uniform
self-oscillations be before the Turing instability of the homogeneous
state \cite{ko:book,ko:ufn89,ko:ufn90,mo2:pre96}. This is a rather
strict requirement for $\al \ll 1$. On the other hand, as was shown in
Sec. IV, the destabilization of the stationary multidomain patterns in
higher dimensions occurs when $\al \sim \ep^{4/3}$, what implies that in
this case $D_\th / D_\et \sim \al^{1/2}$. This condition requires
considerably smaller difference in the diffusion coefficients of the
activator and the inhibitor.

Another interesting implication of our results is that in some
situations it is possible that the oscillatory behavior of the
``homogeneous'' system is actually the consequence of the dynamics of
the underlying multidomain pattern. Indeed, for the reasonable values of
$D_\et \sim 2 \times 10^{-5} {\rm cm}^2 {\rm s}^{-1}$ and $\tau_\et \sim
10 {\rm ~s}$ \cite{epstein95} we would have that the size of an
individual domain is much smaller than $L \sim 1.5 \times 10^{-2} {\rm
  cm}$, so the domains may actually lie beyond the resolution of the
experiment. Notice that in the case $\ep \ll 1$ the oscillatory
instability of the homogeneous state is always accompanied by the Turing
instability \cite{ko:book,ko:ufn89,ko:ufn90}. Numerical simulations show
that Turing patterns may form as a result of the instability of the
homogeneous state even if the homogeneous state is unstable with respect
to the uniform oscillations and may in fact persist for even smaller
$\al$ \cite{mo2:pre96}. So, we suggest that in certain situations it is
the Turing patterns that exhibit the oscillatory behavior, whereas the
system's bulk kinetics has relaxation character.  If this is the case,
one would expect to see the coexistence of the relaxation and
oscillatory kinetics in the system, and multiplicity of the oscillation
modes in a single system with the same parameters. The latter is the
consequence of the fact that the characteristic time and length scales
of the oscillations of the multidomain patterns strongly depend on the
pattern's period which is not uniquely determined by the system's
parameters.

The dynamics of the multidomain patterns discussed in the present paper
was observed by Boissande, Dulos, and De Kepper in the experiments on
CIMA reaction \cite{boissande}. They were actually able to follow the
dynamics of the domains and see synchronization, waves of the
oscillation phase, including spirals on top of the multidomain pattern,
merging of the domains and the coexistence of the domain oscillations
and the uniform oscillations of the homogeneous state (the Hopf-holes,
as they called them). They also emphasized that the waves they observed
are essentially different from those in the Belousov-Zhabotinsky (BZ)
reaction. In the CIMA reaction these are phase waves, as opposed to the
autowaves in the BZ reaction. This is precisely the conclusion of our
analysis.

Synchronously oscillating domains were also observed by Rose {\em et
  al.} in the experiments on the catalytic CO oxidation on the platinum
surface \cite{rose96}. The authors suggested that these oscillations may
be explained by the introduction of global coupling. The results of the
present paper suggest a more natural explanation for this phenomenon as
dynamics of the multidomain pattern. To do this, one has to introduce a
diffusion term into the equation for the inhibitor in the two-variable
reaction-diffusion model of this reaction, which is otherwise known to
have relaxation kinetics \cite{bar94}. This would also seem to explain
the transition from a target pattern into a cellular structure observed
in these experiment. We would also like to mention the observation of
synchronously pulsating cellular flames in the combustion experiments
\cite{gorman94:chaos}.

Haim {\em et al.} observed experimentally a single breathing domain in
the FIS reaction in a circular reactor \cite{haim96}. Their results
agree with the conclusions of Sec. IV concerning the motion of a single
domain. Indeed, they see the supercritical Hopf bifurcation from the
stationary to the breathing domain, growing anharmonicity of the
oscillations for larger amplitudes of the oscillations, and collapse of
the domain for yet larger amplitudes. In this situation Eq. (\ref{rr})
(with $\pi / \sqrt{3}$ replaced by 2) might be used for the quantitative
explanation of these effects. 

The author would like to acknowledge the computational support from the
Center for Computational Science at Boston University.

\bibliography{../main}

\narrowtext

\begin{figure}
  \caption{ The qualitative form of the nullclines of
    Eqs. (\protect\ref{1}) and (\protect\ref{2}).}
  \label{null}
\end{figure}

\begin{figure}
  \caption{The distributions of $\th$ and $\et$ in a stationary
    one-dimensional periodic strata. }
  \label{strata}
\end{figure}

\begin{figure}
  \caption{The parameter regions for oscillations and breakdown of the
    one-dimensional periodic strata. The upper solid line is the
    instability threshold for the multidomain pattern with respect to
    the oscillations. Below the bottom sold line the collective
    oscillations of the strata break down. }
  \label{diag1d} 
\end{figure}

\begin{figure}
  \caption{The dependence $\tau_{1c} (\rb_0)$ from Eq. 
    (\protect\ref{tau1c2}) for different values of $\eb$. }
  \label{tau2d}
\end{figure}

\begin{figure}
  \caption{The nonlinear friction term $f(\rb)$ in
    Eq. (\protect\ref{rr}): (a) $\bar{\ep} = 0.00025, \tau_1 = 0.03$;
    (b) $\bar{\ep} = 0.0075, \tau_1 = 0.175$; (c) $\bar{\ep} = 0.05,
    \tau_1 = 0.025$. Other parameters are $c_1 = 1, c_2 = 0$.  }
  \label{fric}
\end{figure}

\begin{figure}
  \caption{The bifurcation diagram for the two-dimensional stationary
    multidomain pattern for $c_2 = 0$. The numbers 1, 2, and 3 in the
    figure correspond to the bifurcations at $\rb_0 = \rb_{c1}$, $\rb_0
    = \rb_{c2}$, and $\rb_0 = \rb_{c3}$, respectively. The shaded
    regions are those where the instability is not realized ($\tau_{1c}
    < 0$).  }
  \label{reg2}
\end{figure}

\begin{figure}
  \caption{The sweep of the oscillations for given values of $\tau_1$
    (horizontal lines below the solid curve), obtained from the solution
    of Eq. (\protect\ref{bal2}). The parameters used: $\eb = 0.00025$,
    $\rb_0 = 0.25$, $c_1 = 1$, and $c_2 = 0$. }
  \label{fampl}
\end{figure}

\begin{figure}
  \caption{The wave of the phase of the domain oscillations traveling
    from left to right. Distributions of the activator for different
    times.  The parameters used are: $\ep = 0.05, \al = 0.02, A = -0.4$.
    The system is $20 \times 4$. The boundary conditions are periodic.}
  \label{wave}
\end{figure}

\begin{figure}
  \caption{Breakdown of a target wave of the oscillation
    phase. Distributions of the activator for different times. The
    parameters used are: $\ep = 0.05, \al = 0.019, A = -0.2$.  The
    system is $20 \times 20$. The boundary conditions are periodic.}
  \label{chaos}
\end{figure}

\begin{figure}
  \caption{Breakdown of a target wave of the oscillation phase for smaller
    $\al$. Distributions of the activator for different times. The
    parameters used are: $\ep = 0.05, \al = 0.015, A = -0.2$. The system
    is $20 \times 20$. The boundary conditions are neutral.}
  \label{chaos2}
\end{figure}

\begin{figure}
  \caption{Breakdown of the collective domain oscillations into the
    uniform self-oscillations. Distributions of the activator for
    different times. The parameters used are: $\ep = 0.05, \al = 0.015,
    A = -0.1$.  The system is $20 \times 20$. The boundary conditions
    are neutral.}
  \label{unst}
\end{figure}

\begin{figure}
  \caption{The oscillations leading to the interconversion between the
    hot and the cold domains. Distributions of the activator for
    different times. The parameters used are: $\ep = 0.05, \al = 0.015,
    A = 0.1$.  The system is $20 \times 20$. The boundary conditions
    are periodic. }
  \label{flip}
\end{figure}

\end{document}